\documentstyle[12pt]{article}
\newcommand\exa{\nopagebreak \begin{flushleft}\smallskip \nopagebreak
\begin{minipage}[t]{5cm}\sloppy}
\newcommand\exb{\end{minipage}\kern 1cm\begin{minipage}[t]{9cm}\sloppy }
\newcommand\exc{\end{minipage}\kern -3cm \smallskip \end{flushleft}}
\begin{document}
\begin{flushright}
CERN-TH.6564/92  \\
\end{flushright}
\vspace{0.5 cm}

\begin{center} {\Large \bf  Quark confinement, topological
susceptibility and all that in 4 dimensional gluodynamics.}\\
\vspace{1cm}
{\bf A.R.Zhitnitsky}
\footnote{e-mail address is zhitnita@vxcern.cern.ch  }\\
\vspace{1cm}
CERN , Geneva, Switzerland       \\
and\\
Institute of Nuclear Physics, Academy of Sciences of the USSR,
630090 Novosibirsk, USSR.
\end{center}
\vspace{1cm}
\begin{abstract}

We discuss a
few               tightly connected problems, such as the $U(1)$
problem, confinement, the $\theta$ -dependence               within a
framework of the dynamical toron approach. We calculate two fundamental
characteristics  of the theory: the vacuum expectation value (vev)
of the Wilson loop and the topological susceptibility. The analogy
with well known 2+1 dimensional QED which
exhibits confinement   phenomenon
 is also discussed.

\end{abstract}
\vspace {6cm}
\begin{flushleft} CERN-TH.6564/92 \\
July 1992
\end{flushleft}
\vfill\eject
\pagestyle{empty}
\clearpage\mbox{}\clearpage
\setcounter{page}{1}
\pagestyle{plain}

\newpage
{\bf 1.Introduction.}

It is generally believed that the most profound features of the QCD-
confinement and nontrivial $\theta$ dependence , the chiral symmetry
breaking and the absence of the ninth Goldstone boson are tightly
connected. Let me briefly sketch these old arguments.

     We use the standard notations and the $\theta$ vacuum is
introduced to the theory in the following way:
\begin{equation}
\label{1}
\Delta L =\theta Q
\end{equation}
where $Q$ is the topological charge which can be written for
the 4d gluodynamics                                 as follows:
\begin{equation}
\label{2}
Q=\frac{1}{32{\pi}^2}\int d^4x G_{\mu\nu}^a \tilde{G_{\mu\nu}^a}=
\frac{1}{32{\pi}^2}\int d^4x \partial_{\mu}K_{\mu} , \ \ a=1,2,3.\ \
 \mu , \nu =1,2,3,4.
\end{equation}
Here $G_{\mu\nu}^a$ is the field strength tensor.
It is known , the $\theta$ term preserves renormalizability of the theory
but is P and T odd. As it can be seen from (\ref{2}) the $\theta$
term is a full divergence and therefore it is equivalent to certain
boundary conditions .
However as is known, the $\theta$ is the physical parameter of the theory
and
the $\theta$-
dependence of physics is linked to the $U(1)$ problem.
                        \cite{Wit},\cite{Ven} .
Indeed,
  if we believe
that the resolution of the $U(1)$ problem appears within the framework
of these papers  ,we must assume that the
correlator
\begin{equation}
\label{4}
K=i\int d^4x<0|T{\frac{1}{32{\pi}^2} G_{\mu\nu}^a \tilde{G_{\mu\nu}^a}(x)
,\frac{1}{32{\pi}^2} G_{\mu\nu}^a \tilde{G_{\mu\nu}^a}(0)}|0>
\end{equation}
is nonzero in pure Yang Mills theory (YM).
 It means that the vacuum expectation value (vev) of the topological
density
\begin{equation}
\label{5}
<| \frac{1}{32{\pi}^2} G_{\mu\nu}^a \tilde{G_{\mu\nu}^a}|> =
\frac{1}{2}K\theta ,\ \ \ \ \theta \ll 1.
\end{equation}
is also nonzero     . Therefore, the $\theta$ is observable parameter.
The nonzero
vev $<G\tilde{G}> \neq 0$ does imply the CP -violation in physical
transition and leads , for example ,to the mixing of the heavy
quarkonium levels with $J^P=0^+$ and $J^P=0^-$ .
Only dispersion relations are used to translate  $<G\tilde{G}> \neq 0$
into a proof of CP- violation in physical effects.
\cite {Shif}.

So, if we believe that $U(1)$ problem is solved in the framework
of Witten-Veneziano approach \cite{Wit},\cite{Ven}
we automatically get $K\neq 0$ (\ref{4}) and therefore,
the nontrivial $\theta$ dependence.
With introduction of the light quarks (u,d,s) the value of correlator
$K$(\ref{4})      changes \footnote{It is obviously that $K\sim m_q^1$
irrespectively to the number of light flavors \cite{Ven}.},but
 $\theta$ parameter
  is still an experimentally observed   quantity.

The next link comes      from the analysis of the Ward Identities
(WI) which imply \cite{Crew}
that the $U(1)$ problem is linked to the chiral symmetry breaking
phenomenon. Indeed, the relevant WI ( in QCD with $N_f$ flavors)
takes the form:
\begin{equation}
\label{6b}
K_{QCD}=\frac{m}{N_f}<\bar{\psi}\psi>-\frac{m^2}{N_f^2}\int dx
\bar{\psi}\gamma_{5}\psi (x),
\bar{\psi}\gamma_{5}\psi (0)
\end{equation}
and if we have the dynamical solution of the $U(1)$ problem
(by other words,we are in position to calculate the singlet correlator
in (\ref{6b}) to show its smallness in the chiral limit) we should explain
the relation between                $<\bar{\psi}\psi>$ condensate
 ,responsible for the
chiral symmetry breaking and topological susceptibility , responsible
for the solution $U(1)$ problem.

Besides that , in a similar way one can check that   nontrivial
$\theta$ dependence in pure YM theory is through $\theta/N$ at large $N$
( in particular,$ <\tilde{G}G>\sim \sin(\frac{\theta}{N})$ )\cite{Ven}.
 Such a function
can be periodic in $\theta$ with period $2\pi$ only if there are many
vacuum states for given values of $\theta$. These vacua should not be
 degenerated  due to the vacua transitions , however the trace of the
 enlargement number of the vacuum states have to be seen in the course
of a dynamical solution of the $U(1)$ problem. Therefore , we have a link
between the $U(1)$ problem and an $ additional~(discrete)~ classification
$ of the vacuum states in the theory apart from the standard $\theta$
classification.

    Another link , between $U(1)$ problem and confinement phenomenon ,
can be understood from the analysis of the effective lagrangian
describing the low-energy spectrum and dynamics of the pseudoscalar
nonet in the large $N$ limit \cite{Wit3},\cite{Ven2}.
The most important assumption which has been made in the deriviation
of the corresponding effective lagrangian was $ confinement$.
As it is known the obtaining lagrangian perfectly
 describes all properties
mentioned above             . In particular this lagrangian reproduces
the correct $\theta/N$ dependence in YM theory.

Indeed, gluodynamics can be understood as a QCD with very large
quark's mass. In this limit effective lagrangian was    founded in
\cite{Wit3} and it turns out that the number of vacua
is of order $N$ at $N\rightarrow\infty$. This fact actually
is coded in the effective lagrangian containing the
multi-branched logarithm $\log\det (U)$.
In the Veneziano approach \cite{Ven} the same fact can be seen
from the formula for multiple derivation of  the topological
density $Q$ with respect to $\theta$ at $\theta =0$.
\begin{equation}
\label{a}
\frac{\partial^{2n-1}}{\partial\theta^{2n-1}}<Q(x)>\sim
(\frac{1}{N})^{2n-1} \ \ ,n=1,2...
\end{equation}

Therefore,the main idea of this sketch is as follows. All problems under
consideration are tightly connected.Thus , any selfconsistent
dynamical solution of one of them should be necessarily accompanied
by the resolution of the rest problems within same approach.

The purpose of this letter is to demonstrate that such links indeed
take place within framework
of the dynamical toron \footnote{ We keep the term "toron", introduced
in ref.\cite{Hoo1}. By this means we emphasize the fact that the
considering solution minimizes the action and carries the topological
charge $Q=1/2$,i.e. it possesses all the characteristics ascribed
to the standard toron \cite {Hoo1}. However I should note from the
very beginning of this paper, that our solution has nothing to do with the
standard toron and it is formulated in principle in another way than in
ref.\cite {Hoo1}. The keeping of this term   has a historical origin.}
 approach
which was discussed early in context
of different field theories, see ref.\cite{Zhi1} and references therein.
I would like to recall that in all known
 cases, the
toron calculations give, at least, the  selfconsistent results. Thus, one
may expect that the analogous situation should takes place in the theory
under consideration,i.e. in 4 dimensional gluodynamics with $SU(2)$ gauge
group.

 We discuss the statistical ensemble of quasiparticles
which, presumably
  \cite{Zhit}, describes the grand partition function of the
4d YM theory and which possesses  strong quasiparticle interaction.
It was shown that this ensemble describes the system with nontrivial
$\theta$- dependence irrespectively to the strength of  quasiparticle
interaction .
It should be noted,that the coexistence of the strong quasiparticle
interaction and the nontrivial $\theta$ dependence is the nice feature of YM
 theory .                            3d QED   also possesses the long
range quasiparticle interaction and confinement phenomenon
\cite{Pol2}. In this simple model
    physics becomes $\theta$ independent just because of the
strong quasiparticle interaction \cite{Verg}.
 Crucial point
                                        is a very nontrivial algebraic
structure of the quasiparticle interaction in YM theory,
in comparision with analogous calculation of ref.\cite{Verg}
in Polyakov's model.

We will discuss the main assumptions which have been made in the
description \cite{Zhit} of this statistical ensemble a bit later.Now
I would like to cite the result of  calculation of the topological
density as a simplest application of the formulae obtained in ref.
\cite{Zhit}:
\begin{equation}
\label{7b}
<\tilde{G}G>=i2{\Lambda}^4\sin(\frac{\theta}{2}),\ \ \ \
-\pi\leq\theta\leq\pi ,
\end{equation}
where the $\Lambda$ is some renormalization invariant
calculable combination
\footnote{Formula (\ref{7b}) is written in Euclidean space because
its calculation based on the self dual solution defined in this space.
In contrary, the formulae (\ref{4},\ref{5}) is in Minkowski space.}

The nonzero value for  topological density and its $\theta$ dependence
are in agreement with main assumption of Witten-Veneziano approach
\cite{Wit},\cite{Ven}.
Indeed, the formula (\ref{7b}) implies that the topological
susceptibility which is nothing but derivative  of the topological
density with respect to $\theta$, is nonzero. Besides that, this
formula is in agreement with Veneziano expression (\ref{a})
for the multiple deriviation with respect to $\theta$.

As it was argued above, the dynamical calculation of the topological density
should be accompanied by the resolution of the rest problems.
One such link was established already in the previous papers \cite{Zhit},
\cite{Zhi2}. Namely,it was demonstrated that the calculation of
$<\tilde{G}G>$ is accompanied by appearing of the
additional  quantum number             (apart from the standard
parameter $\theta$) classifying a vacuum states.

In this letter we will consider  the less trivial applications
of the formulae describing the grand partition function
of YM theory , ref.\cite{Zhit}. Namely, we will calculate the vev of
the Wilson loop to demonstrate the confinement in this theory.
This is the main result of the work.

Besides that,
we      find the topological susceptibility by explicit calculation
(  and not by differentiating
$<\tilde{G}G>$ with respect to $\theta$) in order
 to check the selfconsistency
of the approach. In such calculation we are able to keep the nonleading
term (as $k\rightarrow 0$) in the correlator for the susceptibility.
It turns out ,that this term is of order ($k^4$) and not of order ($k^2$)
as it could be expected naively. Such behavior for the correlator
is the direct result of the strong quasiparicle interaction
and might have a phenomenological  consequences.For example,
    the proton "spin" problem
can be reduced to the problem of calculation of the first moment
 of the topological susceptibility \cite{Ven3}.

Before we proceed to the detail consideration of the Wilson loop
and topological susceptibility let me briefly formulate the basic
assumptions of the toron approach.

i) I allow the configurations with fractional topological charge
(one half for $SU(2)$ group) in the definition of the functional
integral. It means that a multivalued functions appear in the functional
 integral. However, the main physical requirement is - all gauge
invariant values must be singlevalued . Thus , the different cuts
          accompany   the multivalued functions should be unobservable,
i.e. the gauge invariant values   coincide on the upper and on the
lover edges of the cut.

The direct consequence of the such definition of the functional
integral is the appearing of the new quantum number classifying a vacuum
states. Indeed, as soon as we allowed one half topological
charge,the number of the classical vacuum states is increased by the same
factor two in comparision  with a standard classification,
counting only integer winding numbers $|n>$.

Of course,vacuum transitions eliminate this degeneracy.
However the trace of enlargement number of the classical vacuum
states does  not disappear. Vacuum states now classified by two
numbers : $0\leq \theta <2\pi$ and $k=0,1$. Let me repeat that
origin for this is our main assumption that fractional charge
is admitted and therefore the number of classical vacuum
states is multiplied by a factor two.

I have to note that the same situation takes place in the
supersymmetric YM theory, but in this case the vacuum states
are still degenerate after vacuum transitions. The number $k$
in this case just numerates different vacua at the same
$\theta$. However, the gauge classification for the
winding vacua                      is the same
 for both models  (supersymmetric and nonsupersymmetric one)
before vacuum transitions.

ii) The next main point of the toron approach may be formulated as
follows. We hope that in the functional integral of the gluodynamics ,
when the bare charge tends to zero and when we are calculating some long range
correlation function, only certain field configurations ( the toron
of all types) are important. In this case the hopeless problem of
integration over all possible fields is reduced to the problem of summation
over classical toron configurations.I have no proof that the system
of solutions which have been taken into account is a complete system.
But I would like to stress that a lot of problems ( like the $\theta$
dependence , the $U(1)$ problem , the counting of the discrete number
of vacuum states , the confinement , the nonzero value for the vacuum energy
and so on...) can be described in a very simple manner from this
uniform point of view.

                          Both these points are quite nontrivial ones.
However , I would like to convince the reader in the consistency
of these assumptions by considering a more simple models , where
the answers are well known beforehand.
By this reason let me recall in passing that
all calculations, based on the toron solution \cite{Zhi1} demonstrate
its  very nontrivial role
in different field theories.
Most glaringly these effects appear in supersymmetric variants of a theory.
In particular, in the
 supersymmetric  $CP^{N-1}$-theories, the torons (point
defects) can ensure a nonvanishing value for the
$<\bar{\psi}{\psi}>\sim\exp(2i\pi k/N+i\theta/N)$ with right
$\theta$-dependence. Such behavior is in agreement with the value
of the Witten index which equals $N$\cite{Wit1} and in agreement
with the large $N$-expansion \cite{Wit2}. In analogous way, the chiral
condensates can be obtained for 4d theories: supersymmetric YM (SYM),
supersymmetric QCD (SQCD)( see also calculations
\cite{Coh},\cite{Gom}, based on the standard 't Hooft solution).
 In these cases a lot of various results
are known from  independent consideration (such as the
 dependence of condensates
on parameters $ m,g$; the Konishi anomaly equation and so on...).
\cite{Amat}.  Toron
approach is in agreement with these general results.
The same approach can be used for physically interesting theory of QCD
with $N_f=N_c$. In this case an analogous calculation of $<\bar{\psi}\psi>$
does possible because of cancellation of nonzero modes, like in
supersymmetric theories. For this theory the contribution of the toron
configurations to the chiral condensate has been calculated and is
equal to: $<\bar{\psi}\psi>=-\pi^2\exp(5/12)2^4{\Lambda}^3$ \cite{Zhi1}.
As is well known in any consistent mechanism for chiral breaking a lot
of problems, such as: the $U(1)$-problem, the number of discrete vacuum
states, the $\theta$-puzzle, low energy theorems and so on, must be solved
in an automatic way. We have checked that
 all these properties \cite{Zhi1}  are
                             consistent with    the toron calculation.

{\bf 2.The calculation of the topological susceptibility.}

Let me start by giving a few formulae from ref.\cite{Zhit}.
The grand partition function is given by
\begin{eqnarray}
\label{6}
Z=\sum_{k=0}^{\infty} \frac{{\Lambda}^{4(k_1+k_2)}}{(k_1)!(k_2)!}
\sum_{I_{\alpha},q_{\alpha}}\prod_{i=1}^{k_1+k_2}d^4x_i
exp(-\epsilon_{int.}) ,~~~~~~~~~~~~        \\
\epsilon_{int.}=-\frac{4}{3}\sum_{i>j}q_i I_{i} ln(x_i-x_j)^2
q_j I_{j} +\frac{2}{3}\ln L^2(\sum_{i}q_i I_i)^2, \ \
\Lambda^{4-1/3} =c \frac{M_0^{4-1/3}}{g^2(M_0)}
\exp(-\frac{4{\pi}^2}{g^2(M_0)}) .  \nonumber
\end{eqnarray}
where two different kinds of torons classified by the weight
$I_i$ of fundamental representation of the
$SU(2)$ group and $q_i$ is the sign of the topological charge.
Besides that, in formula (\ref{6}) the value $g^2(M_0)$ is the bare
coupling constant and $M_0$ is ultraviolet regularization, so that
   eq.(\ref{6}) depends on    the renormalization invariant combination
$\Lambda$. As it can be seen from (\ref{6}) the only configurations
     satisfying the neutrality condition
\begin{equation}
\label{6a}
\sum_i q_i I_i=0
\end{equation}
are essential in thermodynamic limit $L \rightarrow \infty$.
In obtaining (\ref{6}) we took into account that the classical contribution
 to $Z$ from $k$ torons is equal to
\begin{equation}
\label{7a}
Z\sim \exp (-\frac{4{\pi}^2}{g^2}k).
\end{equation}
Besides that the factor $d^4x_i$ in eq.(\ref{6}) is due to the 4 translation
coordinates accompany an each toron \footnote{ Let us recall that the
one toron has exactly four zero modes in contrary with instanton
possessing by eight zero modes.}
          and combinatorial factor $k_1!k_2!$ is necessary for avoiding
 double counting for $k_1$ torons and $k_2$ antitorons; lastly, the average
overall configurations ${q},{I}$ is an average over all isotopical directions
and topological charge signs of torons.The constant $c$ in the definition
of $\Lambda$ is  the calculable constant
\begin{equation}
\label{8b}
c=2^5{\pi}^2\exp(-\alpha(1)/2)
\end{equation}
where the coefficient $\alpha(1)$ is tabulated in ref.\cite{Hoo3}.

To compute some vacuum expectation values it is convenient to use
 the correspondence between the grand partition function for the gas (\ref{6})
 and  field theory with Sine-Gordon interaction , as it was done by
Polyakov in ref.\cite{Pol2} for 3d QED. Let us rewrite (\ref{6})
in the form:
\begin{eqnarray}
\label{7}
Z_{\theta}=\int D\vec{{\phi}} exp(-\int d^4x L_{eff.}) ,~~~~~~~~~~~~~~
 \Box\equiv\partial_{\mu}\partial_{\mu}, \\
L_{eff}=1/2(\Box \vec{\phi})^2-\sum_{\vec{I_{\alpha}}}
{\Lambda}^4\exp(i8\pi/\sqrt{3}\vec{I_{\alpha}}\vec{\phi}+i\theta/2)
-\sum_{\vec{I_{\alpha}}}{\Lambda}^4\exp
(-i8\pi/\sqrt{3}\vec{I_{\alpha}}\vec{\phi}-i\theta/2). \nonumber
\end{eqnarray}
In this deriviation it
was used the fact that the logarithm function which
appears in the formula for the interaction (\ref{6}) is the Green
function for the operator $\Box\Box$. After that we can use the method
\cite{Pol2} to express the generating functional in terms of effective field
theory (\ref{7}).

In this effective field theory  the sum over $\vec{I_{\alpha}}$
runs over the $2$ weights of the fundamental representation of
$SU(2)$ group. Note, that the first interaction term is related to
torons and the second one to
     antitorons. Besides that, since we wish to
 discuss the $\theta$ dependence , we also include a term proportional
 to the topological charge density $\frac{\theta}{32{\pi}^2}G_{\mu\nu}
\tilde{G_{\mu\nu}}$ to the starting lagrangian
 and corresponding track from this to the effective
lagrangian(\ref{7}).

The
most important result from ref.\cite{Zhit}
is the nontrivial dependence on $\theta$ of the topological density
and susceptibility.These quantaties       are relevant for the solution
of the $U(1)$ problem :
\begin{equation}
\label{8}
\langle\frac{1}{32{\pi}^2}G_{\mu\nu}
\tilde{G_{\mu\nu}}\rangle \equiv iF(\theta)=i2{\Lambda}^4\sin(\theta/2),
-\pi\leq\theta\leq\pi.
\end{equation}
\begin{equation}
\label{9}
\int d^4x\exp(ikx)\langle\frac{1}{32{\pi}^2}G_{\mu\nu}
\tilde{G_{\mu\nu}}(x),\frac{1}{32{\pi}^2}G_{\mu\nu}
\tilde{G_{\mu\nu}}(0)\rangle_{k\rightarrow 0}\sim
\frac{dF(\theta)}{d\theta}
\sim \frac{1}{2} \cos(\frac{\theta}{2}).
\end{equation}

As was discussed in ref.\cite{Zhi2},
the reason to have the nontrivial $\theta$- dependence (\ref{8},\ref{9})
as well as the strong quasiparticle interaction $\sim \ln(x_i-x_j)^2$
(\ref{6}) in YM theory, is the presence of the nontrivial
algebraic structure $\sim q_i \vec{\mu_{i}} q_j \vec{\mu_{j}} $
in the expression for $\epsilon_{int}$ (\ref{6}). Just this fact
was crucial in the analysis of the 2 dimensional $ CP^{N-1}$ model
also \cite{Zhit}.Such structure for $\epsilon_{int}$ is
in the striking contrast with 2+1 Polyakov's model
\cite{Pol2}, where the interaction energy proportional to the topological
(magnetic) charges of quasiparticles only:
\begin{equation}
\label{10}
\epsilon_{int}\sim \frac{q_i q_j}{|x_i-x_j|}.
\end{equation}
Just this difference leads to the existence
of the nontrivial $\theta$ dependence in 4d YM theory (in spite of the
fact of the strong quasiparticle interaction $\sim \ln(x_i-x_j)^2$)
in contrary with Polyakov's model where physics does not depend
on $\theta$.

We proceed now  to the direct calculation of the topological
susceptibility(\ref{9}).

{}From the eq.(\ref{7}) it is easy to reproduce the formula
(\ref{8}) for the vev of the topological density. To this aim
we differentiate the $Z_{\theta}$ with respect to $\theta$
and put $\vec{\phi}=0$ in the vacuum state.
To compute a correlation functions depending on $x^2$ we have to
introduce 2 different fields $\chi_{\alpha},\alpha=1,2$
in place of the $\vec{\phi}$:
\begin{equation}
\label{11b}
\chi_{\alpha}\equiv\frac{8\pi}{\sqrt{3}}\vec{I_{\alpha}}\vec{\phi}.
\end{equation}
These $\chi_{\alpha}$ fields should be considered as independent
in the course of calculation , because the constraint
\begin{equation}
\label{12b}
\sum\chi_{\alpha}=0 (mod 2\pi)
\end{equation}
should be set to zero only at the end of calculation.
The reason for this is as follows. The constraint (\ref{12b})
is appeared as a manifestation of the neutrality condition
(\ref{6a})in the infinite volume limit of the  functional
 description (\ref{7}) for our ensemble (\ref{6}).
We should be very careful in taking    such limit in the course
of calculation of a different topological characteristics. It can
be argued that this limit should be set at the last stage of
calculation only ( otherwise, we would get some senseless results).
Thus , the constraints (\ref{12b}) as well as (\ref{6a}) will be fixed
at the end of calculation.

With the above consideration taken into account , the topological
susceptibility can be expressed in terms of the new variable
$\chi_{\alpha}$:
\begin{eqnarray}
\label{13b}
\chi_{t}(k)\equiv
\int d^4x\exp(ikx)\langle\frac{1}{32{\pi}^2}G_{\mu\nu}
\tilde{G_{\mu\nu}}(x),\frac{1}{32{\pi}^2}G_{\mu\nu}
\tilde{G_{\mu\nu}}(0)\rangle= \\
-{\Lambda}^8
\int d^4x\exp(ikx)\langle\sin\chi_1(x),\sin\chi_1(0)+
\sin\chi_2(x),\sin\chi_2(0)\rangle \nonumber
\end{eqnarray}
Thus, the problem of calculation of  the topological susceptibility
reduces to the problem of calculation of the following
 correlation function:
\begin{equation}
\label{14b}
\langle M^+_{\alpha}(x_1),M_{\alpha}(x_2)\rangle  ,
\ \ \ \ \  M_{\alpha}(x)\equiv \exp(i\chi_{\alpha}).
\end{equation}
where the operator $M$ is so called disorder operator.

The importance of disorder variables in gauge theories has been emphasized
by 't Hooft \cite{Hoo4}, who has argued that rather than instantons
it is the field configurations with nontrivial $Z_N$ topological charge
that should be considered responsible for the long range confinement of quarks.
Analogous disorder variables have been used in different fields of physics
\cite{Kad}.In the context under consideration this variable was introduced
in 2+1 Polyakov's model in ref.\cite{Snyd} and in two  dimensional $CP^{N-1}$
model in ref.\cite{Zhit}.

A set of operators is now defined as follows. $M$ is an operator that
acts on original $A_{\mu}^a$
 fields by gauge transforming them by $U^{x_0}(x)$; this gauge
 transformation is singular at $x_0$ and has the property that as $x$
encircles $x_0$,$U$ does not return to its original value (
as it happens in the instanton case),
but acquires a $Z_N $ phase ($N=2$ for $SU(2)$ group):
\begin{equation}
\label{15b}
U^{x_0}(\phi=2\pi)=\exp(-i2\pi/N)U^{x_0}(\phi=0),\ \ \
U=\exp(2iI\phi)
\end{equation}
where $\phi$ is an angle variable in the $x-y$ plane and the point
$x_0$ lies at the same plane.
{}From its definition it must be clear that $M(x)$ absorbs one half
topological unit, so we say that $M(x)$ is the annihilation operator
for one point
 toron at $x$ with weight $I$ and $M^+(x)$ is the creation operator
    for one toron.\footnote { Let me note that the small
size instanton ($\rho\rightarrow 0$) at $x_0$ is also can be
created by the singular gauge transormation $U(x)$. However,
in the instanton case $U$ does return to its original
value ( in contrast with (\ref{15b})) as $x$ encircles $x_0$ in the
$xy$ plane.} It should be clear , that $U$ depends on  all $x_{\mu}$
variables , so that $M$ is the annihilation operator for the point
defect. However, at the $x-y$ plane, $U$ depends only on angle variable
$\phi$.
     The singularity of $A_{\mu}=iU^+\partial_{\mu}U$
must be smeared over an infinitesimal region around $x$ as it was done
for the separate toron solution\footnote{ Usually we have in mind
the special kind of regularization which preserves  the self-duality
equation, so that the classical action for this configuration equals
$\frac{8\pi^2}{g^2N}$.}.
    We will not consider the regularization problem in this paper.

Now we would like to demonstrate
that the definition of $M$ as an operator of large gauge transformation
in terms of original fields exactly leads to the expression
(\ref{14b})in terms of the effective field theory (\ref{7}).
To this aim, let us consider the $\epsilon_{int}$ in the formula
(\ref{6}) after the action of the gauge transformation
(\ref{15b}) at point $x_0$. Because this gauge transformation creates an
additional toron at point $x_0$ with isospin $I_0$ in the system of the
other torons placed at $x_i$ with isospins $I_i$ we will obtain an
additional contribution to the $\epsilon_{int}$. Namely,
after action of the operator $M$ we have an additional interaction term between
created toron $I_0$ and torons $I_i$ from the system
\begin{equation}
\label{16b}
\Delta\epsilon_{int}\sim\sum_{i}I_0I_i\ln(x_0-x_i)^2.
\end{equation}

It is easy to understand that this interaction after simply repeating the
deriviation of eq.(\ref{7}),reduces to the following expression in the
 effective field theory:
\begin{eqnarray}
\label{17b}
\langle M(x_0)\rangle=\int D\vec{{\phi}} exp(-\int d^4x L_{eff.})
\exp(i8\pi/\sqrt{3}\vec{I_0}\vec{\phi(x_0)}+i\theta/2),
         \Box\equiv\partial_{\mu}\partial_{\mu}, \\
L_{eff}=1/2(\Box \vec{\phi})^2-\sum_{\vec{I_{\alpha}}=\pm\frac{1}{2}}
{\Lambda}^4\exp(i8\pi/\sqrt{3}\vec{I_{\alpha}}\vec{\phi}+i\theta/2)
-\sum_{\vec{I_{\alpha}}}{\Lambda}^4\exp
(-i8\pi/\sqrt{3}\vec{I_{\alpha}}\vec{\phi}-i\theta/2). \nonumber
\end{eqnarray}
Thus, the operator $M$ under consideration in the effective theory
looks like that
\begin{equation}
\label{18b}
M_{\alpha}(x)=\exp(i8\pi/\sqrt{3}\vec{I_{\alpha}}\vec{\phi(x)}+i\theta/2)=
\exp(i\chi_{\alpha}(x)+i\theta/2),
\end{equation}
where we rewrite eq.(\ref{18b}) in terms of the $\chi$ fields (\ref{11b}).

I would like to emphasize ,
that the expression for the operator of the large gauge transformation
          in terms of the effective variables (\ref{18b}) has
the same form like in another field theories, 2+1 dimensional QED
\cite{Snyd} and  2 dimensional $CP^{N-1}$ models \cite{Zhit}.
As it was argued in the last reference, it is not an accidental
coincidence, but this fact has  a very general origin.

Now we are in position to calculate the correlation function
\begin{equation}
\label{19b}
\int d^4x\exp(ik(x_1-x_2))\langle M(x_1),M^+(x_2)\rangle
\end{equation}
As it was explained in refs.\cite{Pol2},
\cite{Snyd} in the analogous calculations in 3 dimensional QED
we can not neglect terms related to insertions of $M(x)$.
Rather, this functional integral can be estimated by the steepest descent
method. Thus, if we assume that this functional integral is dominated by the
classical field, $\chi$ satisfies
\begin{equation}
\label{20b}
\Box\Box\chi+2{\Lambda}^4 (\frac{4\pi}{\sqrt{3}})^2\sin
\chi=i(\frac{4\pi}
{\sqrt{3}})^2({\delta}^4(x-x_1)-{\delta}^4(x-x_2)).
\end{equation}
The physics suggests linearization is legitimate so we can
substitute $\chi$ instead of $\sin(\chi)$ in the equation (\ref{20b})
and find $\chi_{cl       }(x)$ by means of Fourier transformation.
However, for our purpose an explicit form of $\chi_{cl}(x)$ is
not needed , because we are interesting in the Fourier transformed
form of the correlator (\ref{19b}). Simple calculation gives the following
expression for this correlation function at $k\rightarrow 0$:
\begin{equation}
\label{21b}
\int d^4x\exp(ik(x_1-x_2))\langle M(x_1),M^+(x_2)\rangle
\sim(k^4\frac{3}{64\pi^2}+{\Lambda}^4)^{-1}.
\end{equation}

Now several comments are in order.
First, in comparision  with the well known expression for topological
susceptibility at $k=0$ (\ref{9}) we were able to find the $k$ dependence
at small $k$. It turns out that the $k$ dependence comes in this
formula through $k^4$. Such behavior, from the one hand, is the
direct consequence of the strong logarithmic interaction of
the pseudoparticles (\ref{6}) and from the other hand it implies that
\begin{equation}
\label{22b}
\chi_{t}^{\prime}(k=0)\equiv\frac{\partial}{\partial k^2}
\int d^4x\exp(ikx)\langle\frac{1}{32{\pi}^2}G_{\mu\nu}
\tilde{G_{\mu\nu}}(x),\frac{1}{32{\pi}^2}G_{\mu\nu}
\tilde{G_{\mu\nu}}(0)\rangle _{k=0} \rightarrow 0.
\end{equation}
If such behavior were in QCD theory , this would mean the zero
magnitude for the singlet axial form factor, $G_A(0)$
\cite{Ven3} in agreement with EMC experiment.
However, this is not the case, because we have dealt with the
pure YM theory without quarks and one could expect that the
 dynamical light quarks will change this situation. But
we may expect that such $k^4$ behavior can not disappears
without leaving a trace in the full theory. However , we can only
speculate on this topic now.

As a second remark,
we note that the link between the $\theta$ dependence in the theory,
discrete number of vacuum states there and
confinement phenomenon mentioned in Introduction can be understood
by the following way.

 From the one hand, we have nonzero value
for the vev of the disorder operator $M$ (\ref{18b})
\begin{equation}
\label{23b}
\langle M\rangle \sim \exp(2i\pi k/N+i\theta/N), \ \ \ k=0,1,2... ,N-1.
\end{equation}
In obtaining eq.(\ref{23b}) we took into account that in
    different vacua the fields $\chi_{\alpha}$ take $N$ different
vacuum values $\chi_{\alpha}=2\pi k/N$ , as was discussed
in ref.\cite{Zhit}\footnote{It is easy to understand that
this is the direct consequence of the periodicity of the effective
lagrangian (\ref{7}).Although each of the separate solutions has a
$\theta$ period of $2\pi N$,the overall minimum has a
period $2\pi$ because of jumping from one value to another
at $\theta=\pi$.}.
On the classical level we have $Z_N$ degeneracy , but on the
quantum level, taking into account the toron vacuum transitions,
these vacua have different energy as was the case in the standard
$|\theta>=\sum\exp(in\theta)|n>$ vacuum consideration, but
now there is only a finite number of vacua.The operator
$M$ plays the role analogous to the chiral condensate in QCD.
Different worlds discussed above are labeled by the phase of this
disorder parameter $M$.

{}From the other hand, one could expect that $<M>\neq 0$ just
corresponds to confinement phase. I have no rigorous proof for this,
however , an analogous considerations in 2 dimensional
$CP^(N-1)$ model \cite{Zhit} and in 3 dimensional Polyakov's
model \cite{Snyd}, where vev of the disorder operator
have been calculated ,verify this conjecture.
Moreover, the explicit calculation of the vev of the
Wilson line in 4 dimensional gluodynamics (see the next section),
also demonstrates the nontrivial relation between
nonzero value for $<M>$ and area law for Wilson loop $<W>$.

As a last remark, I would like to note,that
the correlator (\ref{13b}), describing a gluonium states,
can be expressed in terms of the operator $M$. The corresponding
correlation function in the $x$ space
\begin{equation}
\label{24b}
\langle M(x_1),M^+(x_2)\rangle_{(x_1-x_2)\rightarrow \infty}
-\langle M\rangle ^2\sim\exp-|x_1-x_2|
\end{equation}
shows an exponential fall off at large distances.
Thus, one could expect that the operator $M$ can be identified
with the  creation operator of the lowest gluonium state.
It is interesting to note that a solitonic operator
in the different field theories as well as the vertex operator
in the conformal field theories  have the same exponential form.
Thus, the creation operator $M$ which should be highly
nonlocal and nontrivial in terms of the original fields
($A_{\mu}^a $-gluons) has a very simple form in the terms of the
 auxiliary (dual) variables $\chi_{\alpha}$.

{\bf 3. Wilson loop operator}

In order to show directly ( without references to the analogy with
2 and 3 dimensional systems)
 that there are electric strings in the ensemble
of our pseudoparticles (\ref{6}) let us calculate the vev of the
Wilson loop. As was explained above, the torons at large distances
look like singular pure gauge field with definite isotopical
direction and so, the $A_{\mu}^a$ field is abelian at large distances
(in a more detail see  \cite{Zhit}). Thus, the standard quasiclassical
approximation, when we substitute for $A_{\mu}^a$ the corresponding classical
solution, leads ( after simply repeating the deriviation of
 (\ref{7}))  to the following expression for $<W>$ at $\theta=0$
\begin{eqnarray}
\label{25b}
\langle W \rangle=
\langle Tr\exp(\oint_{l}iqA_{\mu}dx_{\mu}) \rangle =~~~~~~~~~~~~~~~~~\\
\int D\vec{ \phi}  exp(-\int d^4x L_{eff.}),\ \
L_{eff}=1/2(\Box \vec{\phi})^2-\sum_{\vec{I_{\alpha}}}
2{\Lambda}^4\cos(8\pi/\sqrt{3}\vec{I_{\alpha}}\vec{\phi}+
\vec{I_{\alpha}}\vec{\Phi}) .    \nonumber
\end{eqnarray}
where term proportional to $\Phi$ is related to Wilson loop
insertion and has the following property : $\Phi(x)$ is equal to the
external charge $2\pi q$ if $x\in S$, Wilson plane, and $\Phi=0$ otherwise.
In this deriviation we took into account that if the toron is in the
$S$ plane, then the integral over $G_{\mu\nu}d\sigma_{\mu\nu}$
is non-zero, and it is equal to zero otherwise (see for a more detail
about toron properties the ref .\cite{Zhit} ).

Since the field $\Phi(x) $ is strong enough , we can not neglect
non-linearities in (\ref{25b}) just like in the calculation
of  the topological susceptibility (see
previous section).  The estimation of this functional integral
can be done as before by the steepest descent method with respect
to $\phi$. The corresponding  field $\phi_{cl}$ is determined
from the classical field equation,
\begin{equation}
\label{26b}
\Box\Box{\chi}^{\prime}+2{\Lambda}^4 (\frac{4\pi}{\sqrt{3}})^2\sin
({\chi}^{\prime})
=2\pi\theta_S(z,t){\delta}'(x){\delta}'(y),
\end{equation}
where ${\chi}^{\prime}\equiv\chi+\vec{\Phi}\vec{I}$, $ {\delta}'(x)
\equiv\frac{d\delta(x)}{dx}$
,$\theta_S(z,t)=1$ if $z,t \in S$ and $\theta_S(z,t)=0$
otherwise.
 The right- hand side of this equation is related to the
Wilson loop insertion,i.e. with function $\Phi(x)$. Indeed,
for the
 Wilson loop placed
 in the $ t-z$ plane we have $\Phi\sim\delta_{x,0}\delta_{y,0}$
and thus , acting by the operator $\Box\Box$ on $\Phi$
we exactly reproduce the right hand side of the eq.(\ref{26b})
( without loss of generality we consider $x>0,y>0$):
\begin{eqnarray}
\label{27b}
\Box\Box\delta_{x,0}\delta_{y,0}\sim\Box\Box (1-\epsilon(x))(1-\epsilon(y))
\sim\Box [{\delta}'(x)(1-\epsilon(y))+{\delta}'(y)(1-\epsilon(x))],\\
\Box\equiv\frac{\partial^2}{\partial x^2}+\frac{\partial^2}{\partial
y^2},\ \ \ x>0, y>0.  \nonumber
\end{eqnarray}
where we have substituted the expression $1-\epsilon(x)$ in place of
$\delta_{x,0}$ and used the following feature of the $\epsilon(x)$
function :$\Box\epsilon(x)\sim{\delta}'(x)$. The next step
of calculation of the r.h.s.(\ref{27b}) is the using of the
Fourier representation for the functions $\epsilon(x)\sim
\int^{\infty}_{0}\frac{\sin kx}{k}dk$ and $\delta(x)\sim
\int^{\infty}_{-\infty}\exp(ikx)dk$
which leads to the desired result:
\begin{eqnarray}
\label{28b}
r.h.s.(\ref{27b})\sim\Box\int d k_x k_x[1-\frac{2}{\pi}\int
d k_y\frac{\sin(k_y y)}{k_y}]
\exp( i k_x x) +(x\rightarrow y)    \\
\sim\int^{\infty}_{-\infty}d k_x k_x\exp(ik_x x)
\int^{\infty}_{0}d k_y k_y\sin(k_y y)+(x\rightarrow y)\sim
{\delta}'(x){\delta}'(y).     \nonumber
\end{eqnarray}

Now we are ready to estimate the functional integral (\ref{25b})
by the steepest descent method.
For this purpose , we substitute $\chi$ instead of $\sin\chi$ in the
eq.(\ref{26b}) and find $\chi_{cl}$ by means of Fourier
transformation just as it was done before in the previous section
for the calculation of the topological susceptibility. We expect
that  the accuracy for such procedure is not very high and the
numerical coefficient given bellow should be considered as an
 estimation of the order of value. With these remarks in mind we
obtain
\begin{equation}
\label{29b}
\langle W\rangle\sim\exp(-\frac{\sqrt{3}\pi}{64\pi}{\Lambda}^2S)
\end{equation}
The result (\ref{29b}) implies that between two fixed charges there
exists an electric string with energy density $\sim {\Lambda}^2$.

Now several comments are in order.First of all, I have to stress
that the unusual kinetic term $(\Box\chi)^2$ in the effective
action (\ref{7})plays a crucial role in this deriviation. From the one
hand ,such kinetic term is the direct consequence of the strong pseudoparticle
interaction (\ref{6}) ,and from the other hand ,just this term
gives a correct expression $\sim{\delta}'(x){\delta}'(y)$
 for the right hand side of eq.(\ref{26b}), leading
to the existence of the solitonic shape solution in
the $x,y$ directions.\footnote{ Let us note that the analogous problem in
3 dimensional QED    looks  as  follow . If Wilson loop is placed in the
$ x y$ plane, the equation , analogous to (\ref{26b}) is essentially
one dimensional \cite{Pol2} and
($\chi_{cl}$)depends only on $z$ . The solitonic
shape for the corresponding solution is due to  the non-linearity
($\sim{\delta}'(z)$)
related to the Wilson loop insertion. This solitonic shape guarantee the
 convergence  of the corresponding integral  $\int dz$,
over perpendicular to Wilson loop direction and leads to the confinement
formula.}
This solitonic shape guarantee the convergence of the corresponding
integral $\int dxdy$ over $x,y$ directions, perpendicular to Wilson
plane $z,t$ and leads to confinement formula (\ref{29b}).
The thickness of the sheet with Wilson loop as boundary is of order
${\Lambda}^{-1}$, plasma screening length, and not $\Delta\rightarrow 0$,
toron size. This is just the reason to get the finite answer for the
string tension (\ref{29b}) in comparision
  with our very rough previous estimation
\cite{Zhit},where back reaction of the Wilson loop insertion has not been
taken into account.

As a second remark ,we note that our explicit calculation of the
$<W>\sim\exp(-S)$ is in the consistency with our conjecture that the
confinement
phase just corresponds to the nonzero value of disorder operator $<M>$
(\ref{23b}). As it was mentioned above, it is also correct in the
2 dimensional $CP^{N-1}$ model \cite{Zhit} and 3 dimensional QED
\cite{Snyd} where an explicit calculations do possible.

{\bf 4.Final remarks}.

The main point of this Letter is that the $ dynamical$ solution of
the one from the following problems : $U(1)$ problem,confinement
problem, multiplicity of the vacuum states $N$, $\theta/N$ dependence,
should be necessarily accompanied by the resolution of the rest problems
within the same approach. We have checked this conjecture in the framework
of the dynamical toron calculation.
I expect that two main assumptions of this approach I am using

i)the multivalued functions are admissible in the definition of
the functional integral;

ii) the only certain field configurations ( torons) are important and
the problem of integration over all possible fields is reduced
to the problem of summation over classical toron configurations,
are consistent  assumptions. I have no proof for this, but
I would like to stress that all problems mentioned above
can be considered from this uniform point of view.

The main quantitative results, which have been obtained can be formulated
as follow:
\begin{equation}
\label{30b}
\chi_{t}(k\rightarrow 0)\sim(\frac{3k^4}{64\pi^2} +\Lambda^4)^{-1}  \\
\end{equation}
\begin{equation}
\label{31b}
<M>\sim\exp(i\theta/2+i\pi k),k=0,1.
\end{equation}
\begin{equation}
\label{32b}
<W>\sim\exp(-\sigma S)
\end{equation}
The first            eq.(\ref{30b})
 shows that the $k$ dependence of the topological
susceptibility comes through $k^4$. Such behavior might be important
for explanation of EMC experiment.

The second eq.(\ref{31b}) explicitly demonstrates that vacua in YM theory
are
classified by the phase of the disorder operator $M$. Apart from standard
parameter $\theta, 0\leq\theta\leq 2\pi$, there is a new label $k=0,1$.
Although each of the separate solutions has a $\theta$ period $4\pi$,
the overall solution has a $\theta$ period of $2\pi$ because
of the jumping from the one value ($k=0$) to another ($k=1$)
 at $\theta=\pi$.
Such behavior is in agreement with large $N$ results, where the
number of vacuum states is equal $N$ and $\theta$ dependence appears
in the form $\theta/N$.
Besides that,$<M>\neq 0$ is in consistency        with
                                 the confinement property of the theory.

 The explicit calculation of the vev
of the Wilson line (\ref{32b}) also confirms this conjecture.

{\bf Acknowledgments}

I am very grateful to G. Veneziano for stimulating
discussions, H.Leutwyler and A.Smilga for  organization of
Moscow - Bern seminar , where this work was presented
 and also to the CERN Theory Division
where this work was done.

\end{document}